# Field-resilient superconducting coplanar waveguide resonators made of Nb, NbTi, and NbTiN

Bongkeon Kim, Chang Geun Yu, Priyanath Mal, and Yong-Joo Doh[*]

Department of Physics and Photon Science, Gwangju Institute of Science and Technology, Gwangju, 61005, Republic of Korea.

Superconducting coplanar waveguide (SCPW) resonators with high internal quality factors ($Q_i$) are essential for quantum information applications, but suffer severe $Q_i$ degradation under magnetic fields due to quasiparticle generation and vortex-induced losses. We fabricated and characterized Nb, NbTi, and NbTiN SCPW resonators with varying film thicknesses under different temperatures and in-plane magnetic fields ($B_\parallel$). Temperature-dependent transmission measurements agree well with Mattis-Bardeen theory, revealing kinetic inductance parameters. Comparative analysis shows that 45-nm-thick NbTi resonators maintain $Q_i = 1.01 \times 10^4$ at $B_\parallel = 0.4$ T, satisfying the dual requirements of high $Q_i$ and strong field resilience. Owing to its moderate kinetic inductance and compatibility with conventional photolithography, NbTi emerges as a practical material platform for field-resilient SCPW resonators and hybrid quantum circuits operating in magnetic environments.

**Corresponding Author**

[*]E-mail: yjdoh@gist.ac.kr

**Keywords:** Superconducting coplanar waveguide resonator, Magnetic field resilience, Quality factor, Resonance frequency.

## 1. Introduction

A superconducting coplanar waveguide (SCPW) consists of a central superconducting microwave transmission line, which is electrically isolated from the surrounding ground planes on an insulating substrate [1]. Resonators patterned in a serpentine geometry are coupled to the central line and designed as either half-wavelength ($\lambda/2$) or quarter-wavelength ($\lambda/4$) structures, depending on whether both ends or only one end are isolated from the ground plane. Owing to their straightforward fabrication, SCPW resonators are widely employed in quantum information technologies, including parametric amplifiers [2], photon detectors [3], quantum memories [4] and hybrid systems that integrate superconducting resonators with spin-based platforms [5-7] or Majorana qubits [8, 9]. For these applications, maintaining a high internal quality factor ($Q_i$) is essential. In particular, spin- or magnetic-field-based hybrid systems require SCPW resonators to sustain high $Q_i$ values under static magnetic fields of approximately 0.1 T to ensure proper operation [5-7].

Maintaining high $Q_i$ under magnetic fields is highly challenging [10], because quasi-particles induced by the field increase dissipation and suppress $Q_i$. In addition, magnetic fields generate Abrikosov vortices in the superconducting film, and their interaction with the microwave signal introduces resistive losses that further lower $Q_i$ [11]. Intrinsic field tolerance can be improved by using superconductors with high critical fields, while reducing film thickness increases the critical field to suppress vortex entry [12, 13]. So far, thin NbN [14] and NbTiN [15] have been employed for magnetic field-resilient SCPW resonators, achieving $Q_i \geq 10^4$ under in-plane fields up to $B_\parallel = 6$ T. However, these materials exhibit large kinetic inductance ($L_k$) due to their long London penetration depth ($\lambda_L$) [16]. Consequently, achieving 50 Ω impedance matching requires sub-micrometer linewidths [15], which are incompatible with conventional UV photolithography and significantly complicate device fabrication.

In this study, we propose NbTi as an attractive alternative to mitigate such fabrication challenges. Owing to its shorter $\lambda_L$ relative to NbN and NbTiN, NbTi exhibits reduced $L_k$ [14, 15, 17], enabling UV photolithography for SCPW resonators while satisfying the impedance-matching condition. We fabricated and characterized SCPW resonators based on Nb, NbTi, and NbTiN films of varying thicknesses, keeping the resonator design identical across devices. A systematic comparison was performed under different temperatures and magnetic fields. At $T$ = 1.8 K, we observed $Q_i$ values between $10^3$ and $10^5$ at zero field, which decreased rapidly with applied field. Notably, 45-nm-thick NbTi SCPW resonators showed promising field resilience, achieving $Q_i$ ($B_\parallel$ = 0.4 T) = 1.01 × $10^4$, exceeding previous reports on NbTi [18] and MoRe [19]. These results highlight NbTi SCPW resonators as promising candidates for spin-based quantum devices [4, 20] and field-resilient quantum circuits such as Majorana-transmon qubits [21, 22].

## 2. Experiments

The Nb, NbTi, and NbTiN SCPW resonators were fabricated on highly resistive intrinsic Si (100) substrates (room-temperature resistivity $\rho$(300 K) = 10 kΩcm), which were covered with a 100-nm-thick layer of thermally grown $SiO_2$. The Nb films were deposited using a pulsed DC magnetron sputtering system with a high-purity (99.9 %) Nb target. NbTi films were deposited using a NbTi alloy target with a 50/50 atomic ratio. During the deposition, the Ar pressure was maintained at 2.0 mTorr and the deposition rate at 0.4 nm/s. NbTiN films were reactively sputtered with a NbTi target in an Ar/$N_2$ (20 sccm/4 sccm) atmosphere, at a deposition rate of 0.3 nm/s. The superconducting transition temperature ($T_c$) and residual resistivity ($\rho_{res}$) for the NbTi and NbTiN films, plotted in Figs. 1a-d, show that $\rho_{res}$ increases and $T_c$ decreases monotonically with decreasing film thickness, consistent with previous findings for $Mo_{0.66}Re_{0.34}$ films [19].

The fabrication process is illustrated schematically in Fig. 2. Following film deposition, the SCPW resonators were patterned using conventional photolithography, and reactive ion etching was performed in an $SF_6/O_2$ (4 sccm/1 sccm) environment. Residual photoresist was removed using mr-Rem 700 remover (Micro resist technology GmbH), followed by rinsing with deionized water and drying using an $N_2$ gas flow [23]. Upon completing fabrication, the resonator chip was Al-wire bonded to a PCB board and then encased in an Au-coated oxygen-free copper box, as shown in Fig. 3a. The assembly was mounted in a home-made RF probe and loaded into a cryogen-free magnet system (Cryogenic Ltd.) equipped with a 9 T superconducting magnet. Measurements of insertion loss spectra ($|S_{21}|$) were performed at a base temperature of $T$ = 1.8 K using a network analyzer (E5071B, Keysight Technologies Inc.) [24]. To avoid Kerr-nonlinearity effects, the microwave power was carefully controlled, maintaining resonance dips at − 40 dBm [2, 25].

Figure 3b shows an optical microscope image of the SCPW resonator chip, which measures 1 cm × 1 cm and includes six hanger-type quarter-wavelength (λ/4) resonators of varying lengths ($l$ = 6.14 – 6.50 mm), all coupled to a central feed line. Figure 3c provides a detailed view, revealing that one end of each λ/4 resonator is inductively coupled to the feedline, while the opposite end is left open to ground. An enlarged image of the feedline in Fig. 3d highlights its width ($s$) and separation ($w$) from the ground planes. The ground plane is patterned with arrays of 5.8 μm × 5.8 μm square holes, designed to pin Abrikosov vortices under applied magnetic fields, thereby preserving a high $Q_i$ value in the resonator.

The resonance frequency ($f_c$) of the SCPW resonators is expressed as $f_c = [(L_g + L_k)C_g)]^{-1/2}/4l$, where $L_k$ is the kinetic inductance per unit length, and $L_g (= (\mu_0/4) K(k')/K(k))$ and $C_g (= 4\varepsilon_0\varepsilon_{eff}K(k)/K(k'))$ denote the geometrical inductance and capacitance per unit length, respectively, which are obtained using the conformal mapping technique. [15] Here, $\mu_0$ is vacuum permeability, $\varepsilon_0$ is vacuum permittivity, and $\varepsilon_{eff}$ is effective dielectric constant. The $K(k \text{ or } k')$ means the complete elliptic integral of the first kind with the parameters $k = s/(s + 2w)$ and $k' = (1 - k^2)^{1/2}$. The values of $L_g$ and $C_g$ are listed in Table 1. The resonators are designed to yield $f_c$ varies between 4.58 GHz and 4.85 GHz, and the characteristic impedance ($Z_0$) is matched to 50 Ω, under the assumption $L_k \ll L_g$.

### 3. Results and discussion

We measured the insertion loss $|S_{21}|$ spectra as a function of applied microwave frequency $f$ for SCPW resonators fabricated from Nb, NbTi and NbTiN films with varying thicknesses. Each spectrum exhibited six resonance dips corresponding to the resonance frequencies $f_c$ of the individual resonators. Figure 4a shows a representative $|S_{21}|$ spectrum for a NbTi SCPW resonators with 45 nm film thickness, where six distinct resonance dips correspond to resonators R1 through R6. To determine the $Q_i$ and $f_c$, we converted $|S_{21}|$ to the transmission coefficient $|S'_{21}|$ using the relationship $|S'_{21}| = 10^{(|S_{21}|-LL)/20}$, accounting for a line loss of $LL$ = -25 dB from the coaxial cables. The normalized transmission spectra of the SCPW resonators were then fitted using the hanger equation [26]:

$$|S'_{21}| = \left|A\left(1 + \alpha'\frac{f-f_c}{f_c}\right)\left(1 - \frac{Q_l e^{i\theta}/|Q_e|}{1+2iQ_l(f-f_c)/f_c}\right)\right| \quad (1)$$

where $A$ represents the feedline transmission amplitude, $\alpha'$ is a background slope parameter, $Q_l$ is the loaded quality factor of the resonator, and $Q_e = |Q_e|e^{i\theta}$ is the complex-valued quality factor. The real part of $Q_e$ defines the coupling quality factor $Q_c$ through $1/Q_c = Re[1/Q_e]$ and the relation between quality factors is given by $1/Q_l = 1/Q_c + 1/Q_i$. Figure 4b shows a typical hanger equation fit for the NbTi(45 nm) R3 resonator, yielding $f_c$ = 3.2605 GHz, $Q_i$ = 1.67 × 10$^4$, $Q_c$ = 1.51 × 10$^3$ and $Q_l$ = 1.38 × 10$^3$. The average fitting parameters obtained from multiple resonators in each sample are summarized in Table 1.

Figure 4c displays the temperature dependence of the NbTi(45 nm)-R6 resonator. With increasing temperature, the resonance frequency progressively shifts to lower values, while the resonance dip becomes broader. Similar temperature-dependent behavior is consistently observed across all Nb, NbTi and NbTiN-based SCPW resonators regardless of film thickness, demonstrating systematic reductions in both $f_c$ and $Q_i$ at higher temperatures. For quantitative analysis, we fitted each resonance curve at different temperatures using the hanger equation of Eq. (1). The analysis shows that the fractional frequency shift, defined as $\delta f_{c,T}/f_c = [f_c(T) - f_c(1.8 \text{ K})]/f_c(1.8 \text{ K})$, is negative for all samples, while $Q_i$ decreases monotonically with increasing temperature, as shown in Figs. 4d and 4e, respectively.

The temperature-dependent behavior of $\delta f_{c,T}/f_c$ and $Q_i$ can be described using the Mattis-Bardeen (M-B) theory [27, 28]:

$$\frac{\delta f_{c,T}}{f_c} = \alpha_k(T)\frac{[\sigma_2(T) - \sigma_2(1.8\ K)]}{2\sigma_2(T)} \quad (2)$$

$$\frac{1}{Q_i(T)} = \alpha_k(T)\frac{[\sigma_1(T) - \sigma_1(1.8\ K)]}{\sigma_2(T)} + \frac{1}{Q_{i,0}} \quad (3),$$

where $\alpha_k = L_k/(L_k + L_g)$ is the kinetic inductance fraction at the base-temperature limit, $Q_{i,0}$ is the temperature-independent quality factor, and $\sigma_1(T)$ and $\sigma_2(T)$ are the real and imaginary components of complex conductivity, $\sigma = \sigma_1 - i\sigma_2$, given by

$$\sigma_1 = \sigma_n \frac{4\Delta(T)}{hf_c} e^{-\frac{\Delta(T)}{k_B T}} \sinh\left(\frac{hf_c}{2k_B T}\right) K_0\left(\frac{hf_c}{2k_B T}\right) \quad (4)$$

$$\sigma_2 = \sigma_n \frac{\pi \Delta(T)}{hf_c} \left[ 1 - 2e^{-\frac{\Delta(T)}{k_B T}} e^{-\frac{hf_c}{2k_B T}} I_0 \left( \frac{hf_c}{2k_B T} \right) \right] \quad (5)$$

Here, $\sigma_n$ is the normal-state conductivity of the film, $\Delta(T)$ is the superconducting energy gap, $h$ is Planck's constant, $k_B$ is Boltzmann's constant, and $I_0(x)$ and $K_0(x)$ are the modified Bessel functions of the first and second kind, respectively. The solid curves in Figs. 4d and 4e were obtained from M-B fits with $\alpha_k$ and $T_c$ as free parameters, as displayed in Fig. 4f. These fits are in good agreement with the experimental data, indicating that the red shift of $f_c$ and the reduction of $Q_i$ in the SCPW resonators at higher temperatures result from increased quasiparticle density caused by thermal breaking of Cooper pairs [19, 29]. This process reduces the superfluid density and increases the London penetration depth ($\lambda_L$) [30].

The fitted $T_c$ parameters in the inset of Fig. 4f exhibit a systematic decrease with reducing film thickness, which is consistent with our measurements in Fig. 1. The kinetic inductance fraction $\alpha_k$, obtained from the M-B fit, increases for thinner films, in agreement with previous results obtained with MoRe resonators [19]. Upon extracting $L_k$ from $\alpha_k$, it is clearly evident that $L_k$ scales inversely with film thickness, as shown in Fig. 4g. This is consistent with the kinetic inductance expression [16]: $L_k = \mu_0 \lambda_L^2 / st$, where $\mu_0$ is the vacuum permeability. Moreover, NbTiN resonators exhibit a substantially larger $L_k$ values compared to Nb or NbTi devices of comparable thickness, far exceeding the geometric inductance $L_g \sim 417$ nH/m reported in Table 1. The enhanced $L_k$ results in a characteristic impedance $Z_0$ exceeding 50 Ω and a lower resonance frequency $f_c$ compared with other resonators of similar geometry (see Fig. 4h and its inset). Thus, the reduced $Q_i$'s observed in NbTiN resonators are attributed to the enhanced $L_k$ values.

Figure 5a shows the insertion-loss spectra of NbTi(45 nm)-R6 resonator measured at $T =$ 1.8 K under varying in-plane magnetic fields ($B_\parallel$). With increasing $B_\parallel$, $f_c$ shifts toward lower values, while the resonance dip amplitude decreases and broadens relative to the zero-field condition. We observed similar red shifts in $f_c$ and reductions in $Q_i$ with increasing $B_\parallel$ across all SCPW resonators examined in this study. By fitting the transmission spectra using Eq. (1), the normalized quality factor, $Q_i(B_\parallel)/Q_i(B_\parallel=0)$, is obtained and presented in Fig. 5b. At low $B_\parallel$, $Q_i$ remains almost constant at its zero-field value. However, beyond a threshold field, $Q_i$ decreases sharply. This behavior is consistent with previous observations in MoRe SCPW resonators [19]. The reduced $Q_i$ is attributed to the field-induced quasiparticle generation and vortex-related dissipation mechanisms [15].

The magnetic field resilience of the SCPW resonators is quantified by the threshold magnetic field ($B_{th}$), defined as the field at which $Q_i(B_\parallel)$ drops to half its zero-field value $Q_i(B_\parallel = 0)$ (dashed line in Fig 5b). Figure 5c shows the averaged $B_{th}$ values as a function of film thickness for each material. NbTiN resonators demonstrate the highest field resilience with $B_{th} = 0.78$ T at 45 nm thickness, while Nb resonators exhibit the lowest resilience at $B_{th} = 0.06$ T for 47 nm film. For NbTi, $B_{th}$ ranges from 0.10 T to 0.61 T, with field resilience increasing as film thickness decreases. Theoretically, the lower critical field for magnetic fields applied parallel to the film surface is given by [31] $B_{c1,\parallel} = (2\Phi_0/\pi t^2) \ln(t/1.07\xi)$, where $\Phi_0 = h/2e$ is the flux quantum, $e$ is the elementary charge, and $\xi$ is the superconducting coherence length. The coherence length can be expressed as [30] $\xi = 0.855\sqrt{h v_F/2\pi^2 \Delta}$, where $v_F$ is the Fermi velocity and $l$ is the mean free path of the film. Using the M-B fit parameters, we obtain $\xi$ = 16.9 nm (Nb), 12.6 nm (NbTi), and 4.7 nm (NbTiN). Figure 5d displays the calculated $B_{c1,\parallel}$ values, in comparison with the experimental $B_{th}$ in Fig. 5c. The deviation observed for Nb films with $t < 50$ nm is attributed to enhanced disorder effects that become prominent when the film thickness approaches the coherence length ($t \sim \xi$).

To evaluate the suitability of SCPW resonators for field-resilient applications, we plot $B_{th}$ and $Q_i(B_\parallel = 0)$ at $T = 1.8$ K in Fig. 5e. This plot reveals distinct trade-offs between field resilience and quality factor across materials. Nb(47 nm) resonators exhibit the highest $Q_i(0) = 6.26 \times 10^4$, but the lowest $B_{th} = 0.06$ T, making them impractical for magnetic field environments. Conversely, NbTiN(45 nm) resonators achieve superior field resilience with $B_{th} = 0.78$ T but suffer from significantly reduced quality factors of $Q_i(0) = 3.62 \times 10^3$, limiting their use in quantum sensing applications, though this limitation could be mitigated through optimized resonator geometry for improved impedance matching. Applying practical performance criteria of $B_{th} \geq 0.1$ T and $Q_i(0) \geq 10^4$ (dashed lines in Fig. 5e) [19], only a few devices meet both requirements: Nb(86 nm), Nb(152 nm), NbTi(45 nm), and NbTi(82 nm). Among these candidates, the 45-nm-thick NbTi resonators provide the optimal balance with $B_{th} = 0.53$ T and $Q_i(0) = 1.52 \times 10^4$, resulting $Q_i(B_\parallel = 0.4$ T$) = 1.01 \times 10^4$. This performance significantly exceeds that of previously reported MoRe(27 nm) resonators, which achieved $B_{th}$ = 0.15 T and $Q_i(0)$ = 1.50 × 10$^4$ [19]. The combination of robust field tolerance and moderate kinetic inductance makes NbTi(45 nm) an optimal candidate for field-resilient quantum circuits.

When the SCPW resonators are misaligned with the applied in-plane magnetic field, a perpendicular field component is introduced. This perpendicular field generates Abrikosov vortices, adding dissipation and suppressing the quality factor $Q_i$. The perpendicular lower critical field is expressed [30] as $B_{c1,\perp} = (\Phi_0 t^2/4\pi\lambda_L^4)ln(\lambda_L^2/t\xi)$ and the $\lambda_L$ is given by $\lambda_L = \sqrt{\hbar\rho_{res}/\mu_0\pi\Delta}$, where $\hbar$ is the reduced Planck constant. Using the M-B fit parameters, we obtain $\lambda_L$ = 298 nm (Nb), 379 nm (NbTi), and 857 nm (NbTiN). Figure 5f displays $B_{c1,\perp}$ as a function of film thickness for Nb, NbTi, and NbTiN. The $B_{c1,\perp}$ values are significantly smaller than $B_{c1,\parallel}$ for the same materials and thicknesses. Furthermore, $B_{c1,\perp}$ decreases as the thickness is reduced. For example, NbTiN(45 nm) exhibits the lowest $B_{c1,\perp} \leq 6$ µT, several orders of magnitude smaller than those of Nb and NbTi. To prevent vortex formation in NbTi(45 nm) SCPW resonators under $B_\parallel$ = 0.1 T, the misalignment angle must remain below $\theta \sim 0.04°$. We note that NbTi yields a larger $B_{c1,\perp}$ than NbTiN due to its smaller $\lambda_L$, providing better robustness against field misalignment compared to NbTiN.

## 4. Conclusion

In this study, we fabricated Nb, NbTi, and NbTiN-based SCPW resonators with varying film thicknesses on Si/SiO$_2$ substrates and systematically investigated their microwave response as a function of temperature and magnetic field. The measured temperature dependence of $Q_i$ and $f_c$ agreed well with Mattis-Bardeen theory, while the extracted kinetic inductance scaled as $L_k \propto 1/t$, confirming its thickness dependence. Under in-plane magnetic fields, the resonators exhibited the expected degradation of $Q_i$. Notably, 45-nm-thick NbTi resonators demonstrated an optimal balance between $Q_i$ and field resilience, maintaining $Q_i > 10^4$ at $B_\parallel$ = 0.4 T. Our observations highlight that NbTi-based SCPW resonators are promising candidates for integration into hybrid quantum information systems operating under magnetic fields.

## Acknowledgement

This work was supported by the NRF of Korea through the Basic Science Research Program (RS-2018-NR030955, RS-2023-00207732, RS-2025-02317602), the ITRC program (IITP-2025-RS-2022-00164799) funded by the Ministry of Science and ICT.

## Author Contribution

B.K. and C.G.Y. performed experiments and analyzed data. P.M. contributed to data analysis. Y.J.D. designed and supervised the research project. The manuscript was written by B.K. and Y.J.D. with input from all authors.

## Data availability

No datasets were generated or analyzed during the current study.

## Declaration of competing interest

The authors declare that they have no known competing financial interests or personal relationships that could have influenced the work reported in this paper.

| No. | material | $N_{res}$ | $t$ (nm) | $s$ (μm) | $w$ (μm) | $L_g$ (nH/m) | $C_g$ (pF/m) | $f_{c, avg}$ (GHz) | $Q_l$ | $Q_i$ | $Q_c$ |
|---|---|---|---|---|---|---|---|---|---|---|---|
| 1 | Nb | 5 | 23 | 12.4 | 6.3 | 418 | 169 | 3.4041 | $1.45 \times 10^3$ | $6.91 \times 10^3$ | $1.84 \times 10^3$ |
| 2 | Nb | 6 | 47 | 12.7 | 5.8 | 403 | 174 | 4.0285 | $2.92 \times 10^3$ | $6.26 \times 10^4$ | $3.07 \times 10^3$ |
| 3 | Nb | 4 | 86 | 11.9 | 6.7 | 432 | 165 | 4.2615 | $4.76 \times 10^3$ | $1.25 \times 10^4$ | $7.69 \times 10^3$ |
| 4 | Nb | 3 | 152 | 11.0 | 7.6 | 457 | 154 | 4.3332 | $2.91 \times 10^3$ | $4.22 \times 10^4$ | $3.12 \times 10^3$ |
| 5 | NbTi | 3 | 26 | 12.0 | 6.6 | 426 | 165 | 2.5909 | $9.09 \times 10^2$ | $1.88 \times 10^3$ | $1.76 \times 10^3$ |
| 6 | NbTi | 6 | 45 | 12.0 | 6.6 | 426 | 166 | 3.2699 | $1.63 \times 10^3$ | $1.52 \times 10^4$ | $1.83 \times 10^3$ |
| 7 | NbTi | 4 | 82 | 11.6 | 7.2 | 441 | 160 | 3.8064 | $2.67 \times 10^3$ | $1.90 \times 10^4$ | $3.10 \times 10^3$ |
| 8 | NbTi | 2 | 132 | 12.0 | 6.8 | 427 | 165 | 4.1008 | $4.46 \times 10^3$ | $2.28 \times 10^4$ | $5.55 \times 10^3$ |
| 9 | NbTiN | 6 | 45 | 12.2 | 6.4 | 422 | 167 | 2.1189 | $1.69 \times 10^3$ | $3.62 \times 10^3$ | $3.17 \times 10^3$ |
| 10 | NbTiN | 2 | 81 | 12.4 | 6.3 | 411 | 172 | 2.5656 | $2.09 \times 10^3$ | $7.95 \times 10^3$ | $2.83 \times 10^3$ |

**Table 1.** Characteristic parameters of SCPW resonators. $N_{res}$ denotes the number of measured resonators per chip. Film thickness ($t$) was determined using atomic force microscopy. Geometrical inductance ($L_g$) and capacitance ($C_g$) were calculated from the central conductor width ($s$) and slot width ($w$) (see text). Resonator properties including resonance frequency ($f_c$), loaded quality factor ($Q_l$), internal quality factor ($Q_i$), and coupling quality factor ($Q_c$) were obtained from spectral fitting (see text)

# Figure captions

**Fig. 1** Electrical transport properties and superconducting characteristics of NbTi and NbTiN thin films. Thickness-dependent resistivity *vs.* temperature curves of (a) NbTi and (b) NbTiN films. Residual resistivity $\rho_{res}$ (squares) and superconducting transition temperature $T_c$ (circles) of (c) NbTi and (d) NbTiN films

**Fig. 2** Schematic illustration of the fabrication process for the SCPW resonator

**Fig. 3** Layout of SCPW resonator chip. (a) Optical images of the chip mounted and Al wire-bonded to oxygen-free copper sample holder. (b) Chip containing a single transmission line coupled to six $\lambda/4$ resonators of varying lengths. (c) Single $\lambda/4$ resonator coupled to the transmission line. (d) Magnified view of the central superconducting line flanked by ground planes with hole arrays, showing the line width (*s*) and gap spacing (*w*)

**Fig. 4** Temperature-dependent microwave response of SCPW resonators. (a) $|S_{21}|$ transmission spectrum of NbTi(45 nm) resonators at $T = 1.8$ K, showing six resonance dips. (b) Magnified view of the resonance dip (circle) from the R3 resonator in (a). The resonance frequency $f_c$ and quality factors are obtained via spectral fitting (solid line). (c) Temperature-dependent spectra of the R6 resonator in (a). Temperature dependence of (d) fractional frequency shift $\delta f_{c,T}/f_c$ and (e) internal quality factor $Q_i$ for Nb, NbTi, and NbTiN resonators (symbols), with Mattis-Bardeen model fits (solid lines). (f) Fitting parameter $\alpha_k$ versus film thickness *t* for Nb (circles), NbTi (triangles), and NbTiN (squares). Inset: M-B fitting parameter $T_c$ as a function of *t*. (g) Kinetic inductance $L_k$ (symbols) versus film thickness with $1/t$ scaling (solid lines). (h) Characteristic impedance $Z_0$ and averaged resonance frequency $f_{c,avg}$ (Inset) as a function of film thickness

**Fig. 5** Magnetic field response of SCPW resonators. (a) Transmission spectrum of the NbTi(45 nm)-R6 resonator under in-plane magnetic fields up to $B_\parallel = 1.4$ T at $T = 1.8$ K. (b) Normalized $Q_i$ as a function of $B_\parallel$. The dashed line indicates $Q_i(B_\parallel)/Q_i(0) = 0.5$, defining the threshold field $B_{th}$. (c) Film thickness dependence of $B_{th}$, with values averaged over resonators on the same chip. (d) Estimated in-plane lower critical field $B_{c1,\parallel}$ as a function of film thickness. (e) Correlation between $B_{th}$ and zero-field $Q_i$, with values averaged over resonators on the same chip. (f) Estimated perpendicular lower critical field $B_{c1,\perp}$ for various materials and film thicknesses

**Fig. 1**

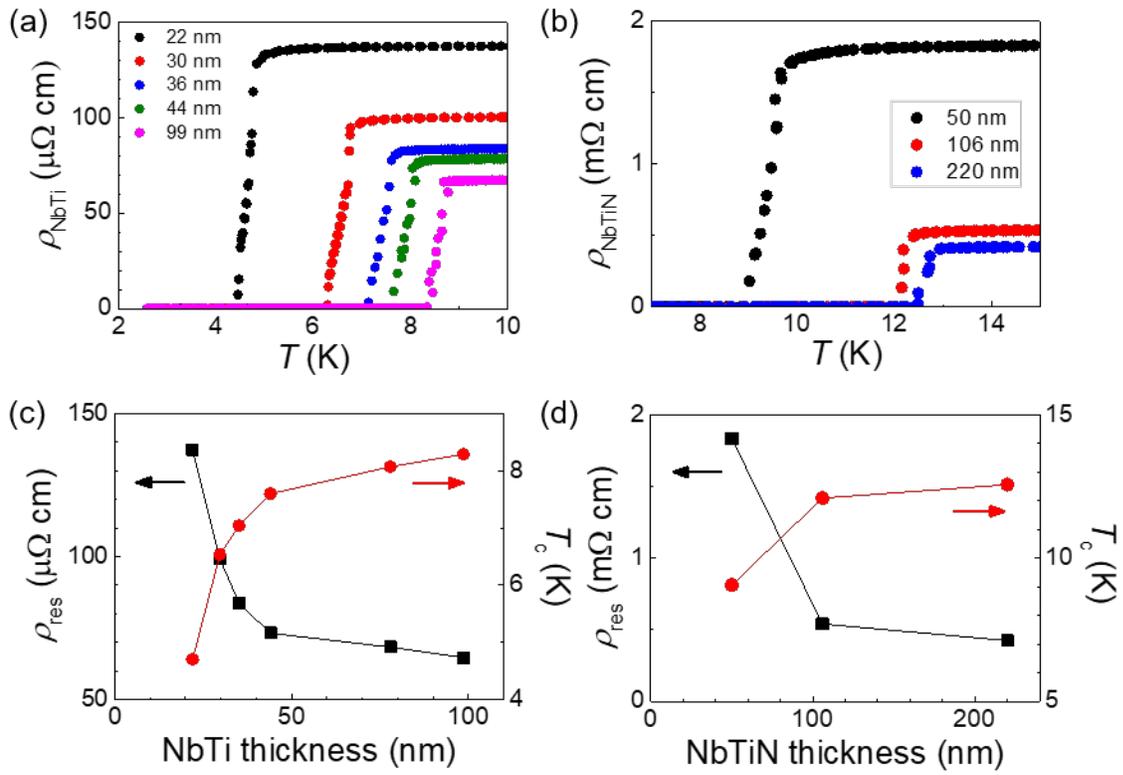

**Fig. 2**

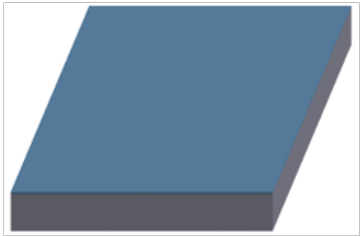
1. Substrate cleaning

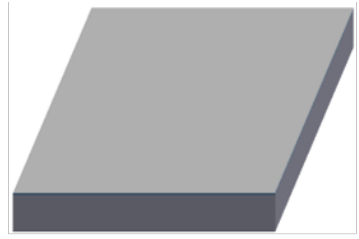
2. Thin film sputtering

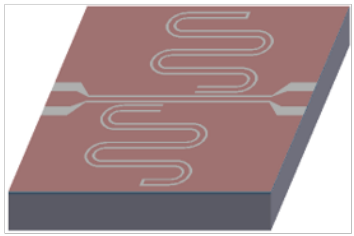
3. Photolithographic patterning

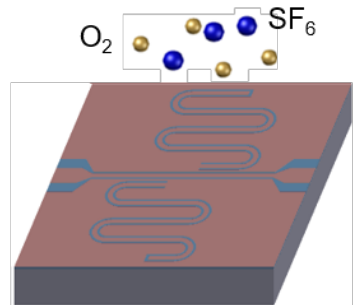
4. Reactive ion etching

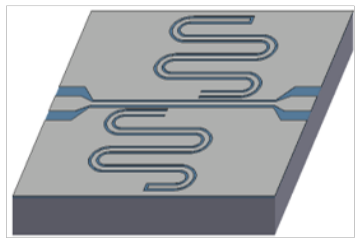
5. Photoresist removal

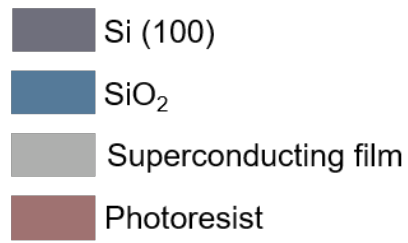

Si (100)
SiO$_2$
Superconducting film
Photoresist

**Fig. 3**

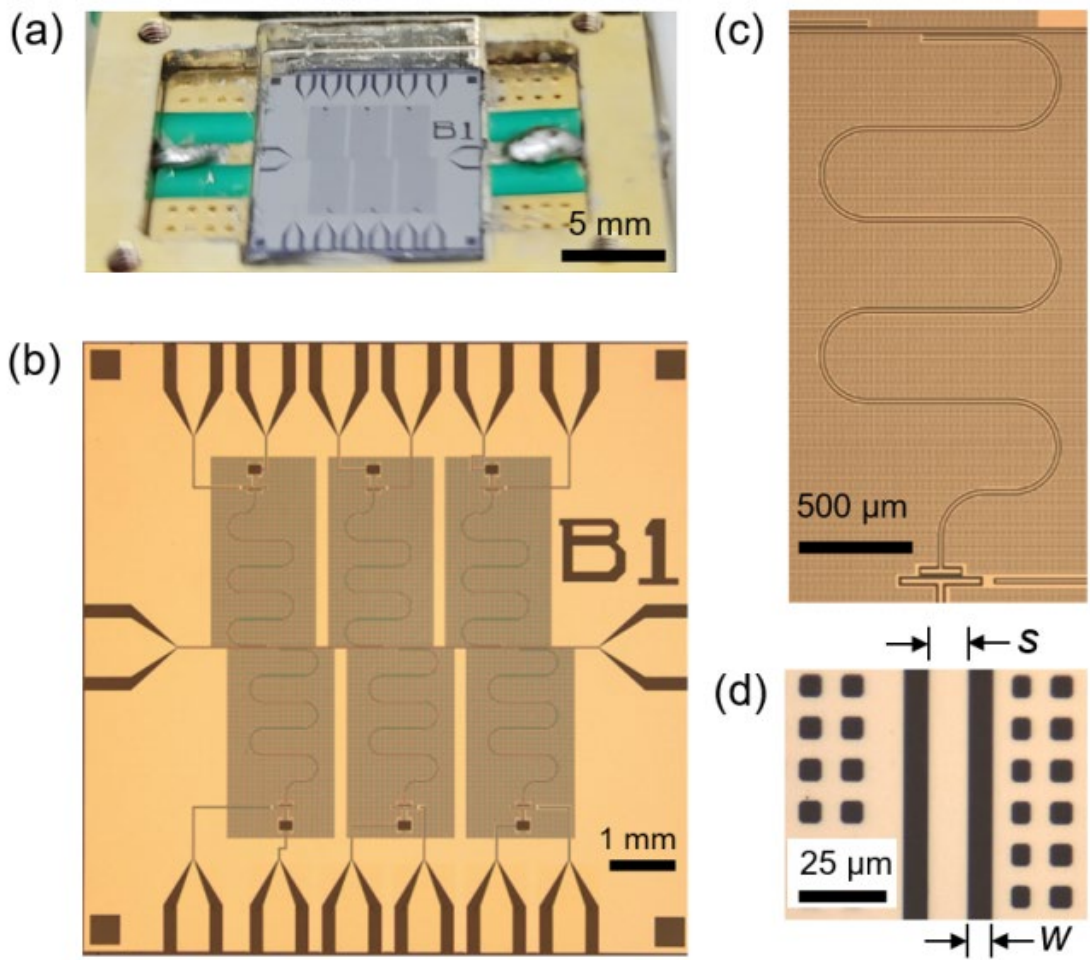

**Fig. 4**

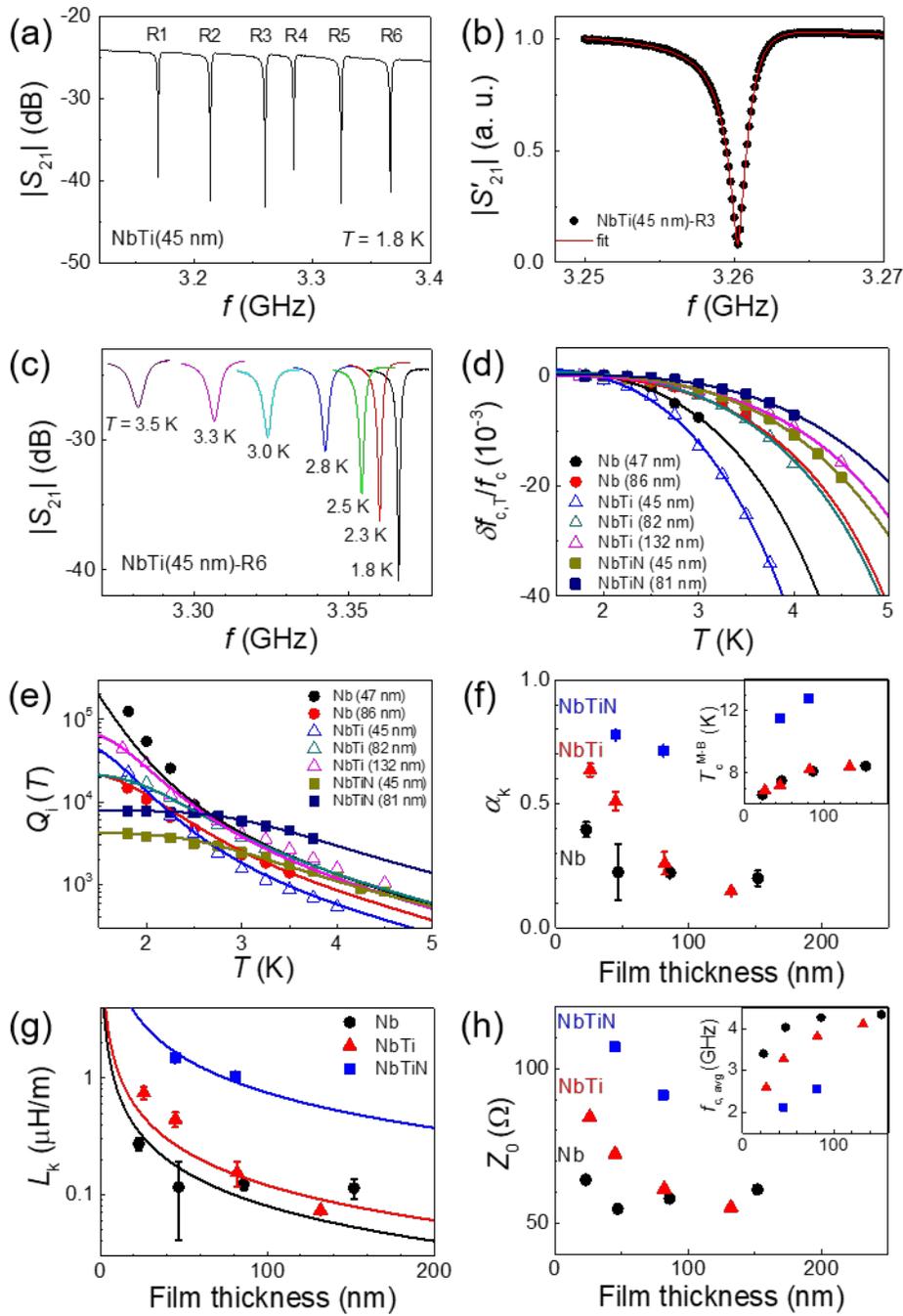

**Fig. 5**

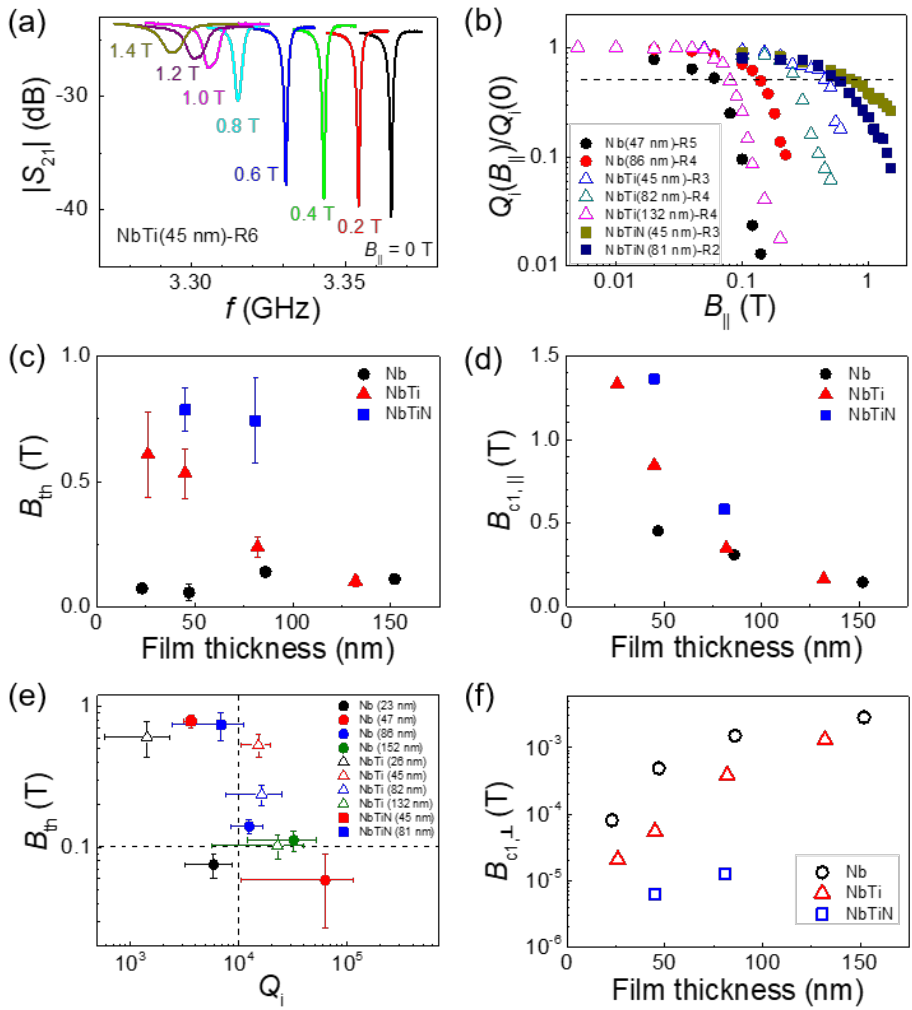